\documentclass[12pt]{amsart}
\usepackage{amsmath}
\usepackage{amsthm}
\usepackage{amsfonts}
\usepackage{graphicx}
\usepackage{amssymb} 
\newenvironment{nouppercase}{
  
  \renewcommand{\uppercasenonmath}[1]{}}{}

\begin{document}

\title[On some new types of membrane solutions]
{On some new types of membrane solutions}
\author[Jens Hoppe]{Jens Hoppe}
\address{Braunschweig University, Germany}
\email{jens.r.hoppe@gmail.com}

\begin{abstract}
New classes of exact M(em)brane solutions in $M+2$ dimensional Minkowski space are presented
(some describing non-trivial topology changes, while others explicitly avoid finite-time singularity formation)
\end{abstract}

\begin{nouppercase}
\maketitle
\end{nouppercase}
\thispagestyle{empty}
\noindent
Exact solutions to non-linear PDE's are usually easy to check, but difficult to find. Concerning membrane solutions in 4 dimensional space-time decades passed between Dirac's spherically symmetric solution \cite{1} and (cp.\cite{2})  
\begin{equation}\label{eq1} 
\begin{array}{l}
(t^2+x^2+y^2-z^2)(t+z)^2  = C (<0) \\[0.15cm]
(t^2-x^2-y^2-z^2)  = C(t+z)^6 (>0). 
\end{array}
\end{equation}
When wanting to find solutions via the level set method in $D$--dimensional Minkowski space, $\mathbb{R}^{1,D-1}$, one has to find functions $u(x^{\mu = 0,1,...,D-1})$ that, on the set defined by $u = 0$ satisfy (see e.g. \cite{BH93}) 
\begin{equation}\label{eq2} 
(\eta^{\mu\nu}\eta^{\rho\lambda} -  \eta^{\mu\rho}\mu^{\nu\lambda}) u_{\mu}u_{\nu}u_{\rho\lambda} = 0
\end{equation}
where $u_{\mu} := \frac{\partial u}{\partial x^{\mu}}$, $u_{\rho\lambda} := \frac{\partial^2 u}{\partial x^{\rho}\partial x^{\lambda}}$, $\eta^{\mu\nu} := \text{diag} (1,-1, \ldots -1)$.\\[0.25cm]
(\ref{eq1})$_{C>0}$ e.g. could be derived by inserting the Ansatz $u(x) = -x^{\mu}\eta_{\mu\nu}x^{\nu} + g(\alpha\cdot x)$, $\alpha^2 = 0$, into (\ref{eq2}), and then noting that, on $u = 0$, the resulting ODE for $g$ actually becomes linear, giving $g(v) = C(t+x)^{2(D-1)} + B(t+x)$. Similarly it is not difficult to verify that 
\begin{equation}\label{eq3} 
u(x^{\mu})  =  -x^{\mu}\eta_{\mu\nu}x^{\nu} + g(\alpha\cdot x) + h(\beta \cdot x), \quad \alpha \cdot \beta = 0
\end{equation}   
will satisfy (\ref{eq2}) on $u=0$ if the following 3 ODE's are satisfied: 
\begin{equation}\label{eq4} 
\beta^2 h''(\alpha^2 g'^2 - 4vg' + 4g) + \alpha^2 g''(\beta^2 h'^2 - 4wh' + 4h)  = 0,
\end{equation} 
\begin{equation}\label{eq5} 
-\alpha^2 g'^2 + 4vg' + 2v^2g'' -4g + D[\alpha^2g'^2-4vg' +4g] -2\alpha^2 g g''  = 0
\end{equation} 
and the same for $w$, $h(w)$ instead of $v$, $g(v)$. \\
Below I would like to discuss the case $\alpha^2 \neq 0 = \beta^2$ in some detail (in particular for $D = 3$ and 4, i.e. strings and membranes resp.).
(\ref{eq4})$_{\beta^2 = 0}$ implies $h(w) = B \cdot w$ (which for $\beta \cdot \alpha = 0$ can always be added, as arising from a constant space-time translation from $B = 0$). $g(v) = f \frac{(\alpha \cdot x)}{\alpha^2}$ on the other hand gives the (a priori still not entirely simple looking) ODE 
\begin{equation}\label{eq6} 
(D-1)[f'^2 - 4vf' + 4f] = 2f''(f-v^2),
\end{equation} 
which however not only has the property that $f(v) = v^2$ makes both sides separately vanish, but also reduces to the ``significantly simpler'' ODE
\begin{equation}\label{eq7} 
2\varepsilon''\varepsilon = (D-1)\varepsilon'^2 + 4(D-2)\varepsilon 
\end{equation} 
upon $f(v) = v^2 + \varepsilon$. An important subtlety\footnote{which could easily be overlooked; a `separation of variables' point of view concerning the Euclidean case has been discussed in \cite{JH19},p.16 and in an unpublished manuscript with J.Choe and V. Tkatjev; separation of variables in the Minkowski case was first discussed in \cite{JH95}, for $D=4$.} is that, depending on the range of $\varepsilon$, resp. $\frac{\varepsilon'^2}{\varepsilon}$, (\ref{eq7}) can be reduced to 3 (qualitatively very) {\it different} ODE's, namely  
\begin{align}
\varepsilon'^2 & = \gamma^2 \varepsilon^{D-1} - 4\varepsilon  & (\frac{\varepsilon'^2}{\varepsilon} > 0) \\
\varepsilon'^2 & = 4(-\varepsilon) - \gamma^2 (-\varepsilon)^{D-1}  & (-4 < \frac{\varepsilon'^2}{\varepsilon} < 0) \tag{$8'$}\\
\varepsilon'^2 & = 4(-\varepsilon) + \gamma^2 (-\varepsilon)^{D-1}  & (\frac{\varepsilon'^2}{\varepsilon} < -4). \tag{$8''$}
\end{align}
While (8) and ($8''$) give time-like minimal surfaces, ($8'$) has the very interesting property that (as I will show in detail for $D=3,4$) it describes topology {\it changing} (space-like) M-brane motions, namely: infinitely many compact, spherical, disconnected M-branes (for $t^2 <1$) becoming 2 (for $D = 3$) disconnected infinitely extended strings resp. (for $D\geqslant 4$) one infinitely extended Membrane for $t^2 >1$.
In each of the cases the constants can be scaled to certain convenient values (note that $u(x)=0$ and $u(\lambda x) = 0$ describe trivially related minimal surfaces), namely (for $\varepsilon > 0$)
\begin{equation}\label{eq9} 
h'^2 = 4(h^2 - h) \qquad \text{for} \; D= 3
\end{equation} 
\begin{equation}\label{eq10} 
h'^2 = 4h(h^2-1) \quad \text{for} \; D= 4.
\end{equation} 
Writing $h =H^2$ in (\ref{eq9}) gives $H'^2 = H^2-1$, which is solved by $H = \cosh$, and 
\begin{equation}\label{eq11} 
u(t,x,z) = -t^2+x^2-\cosh^2 z = 0
\end{equation} 
(note that the effect of $v^2 = \alpha_{\mu}\alpha_{\nu}x^{\mu}x^{\nu}$ for $\alpha_{\mu} = (0, \ldots , 0, 1)$, $\alpha^2 =-1$, is to remove $z^2$ from $x^{\mu}x_{\mu}$) gives
\begin{equation}\label{eq12} 
x^2(t,z) = t^2 + \cosh^2 z,
\end{equation} 
i.e. for each $t$, two infinitely extended strings that resemble (and for $t=0$ {\it are}) the catenary curve. As 
\begin{equation}\label{eq13} 
\begin{split}
(\frac{1}{2}\partial_{\mu}u)^2 & \;\;\:= t^2 - x^2 - \cosh^2 \; (\sinh^2 = \cosh^2 -1)\\
 & \underset{\text{on}\:u=0}{=} -\cosh^4 < 0
\end{split}
\end{equation} 
the entire world-sheet (consisting of 2 disconnected pieces) is time-like, i.e. all points of the 2 infinitely extended strings move with velocity $< 1$. 
In the case of $(8')_{D=3}$ one gets (now with $h: - \varepsilon = H^2$) the ODE
\[
H'^2 = 1-H^2 \tag{$9'$}
\]
which is solved by $H = \cos(v-v_0)$, and for $(8'')_{D=3}$
\[
H'^2 = 1+H^2 \tag{$9''$}
\]
which is solved by $H = \sinh(v-v_0)$.\\
In the former case one gets
\[
u = -t^2+x^2 + \cos^2(v-v_0) = 0 \tag{$11'$}
\]
\[
x^2(t,z) = t^2 - \cos^2 z; \tag{$12'$}
\]
note that the appearance of $\cos z$ (i.e. the same function of a {\it spatial} variable than the radius is as a function of time, when starting with a circular string at rest) is no coincidence: had one chosen $\alpha = (1,0, \ldots,0)$ instead of $(0, \ldots, 0, 1)$ one would have eliminated the $t^2$ term in $x^{\mu}x_{\mu}$ and obtained $x^2 + z^2 = \cos^2 t = r^2(t)$
(here $\alpha^2 = +1$, but $\varepsilon < 0$ giving a $-$ sign to obtain $-x^{\mu}x_{\mu} + t^2 - \cos^2 t$); ($11'$) implies
\[
(\frac{1}{2}\partial_{\mu}u)^2 = t^2-x^2 -\cos^2(1-\cos^2) \underset{u=0}{=} +\cos^4 \geqslant 0        \tag{$13'$}
\]
(i.e. space-like for $z \neq (2n+1)\frac{\pi}{2}$, light-like for $z = (2n+1)\frac{\pi}{2}$). In the latter case, on the other hand
\[
u = -t^2+x^2 + \sinh^2 z =0 \tag{$11''$}
\]
\[
x^2(t,z) = t^2 - \sinh^2 z  \tag{$12''$};
\]
\[
(\frac{1}{2}\partial_{\mu}u)^2 = t^2 - x^2 - \sinh^2(\sinh^2+1) \leqslant 0   \tag{$13''$}
\]
(i.e. time-like for the $x = \pm t$ lines at $z=0$, and the only singularity of the world-sheet being at the origin, $t=x=z=0$).
$(12'')$ shows that for every $t(\neq 0)$ the curve closes (at $x =0$), and the solution is (for $t>0$) a growing (shape-wise flattening) convex closed string centered at $x=z=0$. $(12')$ on the other hand shows a topology change at $t^2 = 1$, as for $t^2 < 1$ $x^2= 0$ can be reached for some (infinitely many) value(s) of $z$, hence describing (for $t^2 <1$) infinitely many convex closed curves (centered at odd multiples of $\frac{\pi}{2}$), while for $t^2 > 1$ describing 2 smooth infinitely extended $(z \in (-\infty, +\infty))$ strings, with the out-most (concerning $x$) points moving (for $t>1$) with velocity $+1$ to the right resp. $-1$ to the left, while all other points moving faster (such that the strings `flatten'). At $t^2 =1$ the shape is singular at heights $z$ being a multiple of $\pi$. While, given the vast literature on strings (time-like minimal surfaces), it is likely that some of what is presented above has been noticed before, it certainly is a good preparation for the now following discussion of membrane motions, where the shapes involve elliptic, rather than trigonometric and hyperbolic functions.\\
So, now for $D=4$: with $\varepsilon(v) = \frac{2}{|\gamma|}Q(\sqrt{\frac{|\gamma|}{2}}v)$ the 3 cases in (8)... become
\begin{equation}\label{eq14} 
Q'^2 = 4Q(Q^2-1) \qquad (\frac{Q'^2}{Q}\geqslant 0)
\end{equation} 
whose solution is the Weierstrass $\wp$-function $P$(with $g_3 =0$, $g_2 =4$) with $P(\omega) = 1$, $P(z) \geqslant 1$ ($z \in (0,2\omega)$) periodic with period $2\omega$ and diverging at even multiples of the half-period $\omega$.
\[
Q'^2 = 4(-Q)(1-Q^2) \qquad (0 \geqslant \frac{Q'^2}{Q} \geqslant -4) \tag{$14'$}
\] 
(with $Q = \frac{-1}{\wp} =: p \in [-1,0]$) and 
\[
Q'^2 = 4(-Q)(Q^2+1) \qquad (\frac{Q'^2}{Q} \leqslant -4) \tag{$14''$}
\] 
which upon $Q = -H^2$ becomes $H'^2 = H^4 +1$ (which with $H$ also has $\frac{1}{H}$ as a solution of the same ODE; the solutions are called hyperbolic lemniscate sine ($slh$) resp. cosine ($clh$) function; just as the ODE for the Weierstrass $\wp$-function is invariant under $Q \in[1,+\infty] \rightarrow \frac{-1}{Q} \in [-1,0]$; note that $H(z)$ has to become infinite for finite $z$, just as the solution of $H'^2 = H^4 +1 + 2H^2$ which is $\tan(z-z_0)$).\\
The corresponding minimal hyper-surfaces (leaving out some free constants) are then described by, firstly,
\begin{equation}\label{eq15} 
u(x^{\mu}) = -t^2 + x^2 + y^2 - P(z) = 0,
\end{equation}
i.e.
\begin{equation}\label{eq16} 
r^2(t,z)(= x^2+y^2) = t^2 + P(z), 
\end{equation}
and
\begin{equation}\label{eq17} 
(\frac{1}{2}\partial_{\mu}u)^2 = t^2 - r^2 - \frac{1}{4}P'^2 = t^2-r^2-(P^3-P) \underset{u=0}{\cong} - P^3 < 0;
\end{equation}
which, in contrast to (\ref{eq12}), describes the time-evolution of an axially symmetric surface, whose contour is {\it not one} curve (when $z \in (-\infty, +\infty)$) but consisting of infinitely many (geometrically identical) pieces `squashed' between even multiples of the half-period $\omega = \frac{1}{\sqrt{2}}K(\frac{1}{\sqrt{2}}) = \frac{1}{4\sqrt{2\pi}}(\Gamma\frac{1}{4})^2$ (cp.\cite{JH95}), each having $r = \sqrt{t^2 +1}$ as its closest approach to the $z$-axis, moving with velocity $\frac{t}{\sqrt{t^2+1}}$, and all other points on the 2 dimensional surface(s) moving with smaller velocity (outward for $t>0$, inward for $t<0$).\\
Secondly, 
\[
u(x^{\mu}) = -t^2 + x^2 + y^2 + (sl(z))^2 = 0 \tag{$15'$}
\] 
(where $sl(z)$, called the Lemniscate sine function is a Jacobi-elliptic function of period $4\omega$, satisfying $sl'^2 = 1 -sl^4$, $sl(0) = 0$; in ($15'$) one could of course just as well have taken the Lemniscate cosine function $cl(z)$, $cl'^2 = 1 - cl^4$, $cl(0)=1$, to improve the analogy with ($11'$)),
\[
r^2(t,z) = t^2 - (sl(z))^2, \tag{$16'$}
\]
and
\begin{align*}
(\frac{1}{2}\partial_{\mu}u)^2  & = t^2 - x^2 - y^2 - (sl(z))^2(sl'(z))^2  \tag{$17'$} \\ 
    & = t^2 - x^2 - y^2 - (sl(z)^2)(1-sl^4(z)) \cong (sl(z))^6 \geqslant 0; 
\end{align*}
for $t^2 < 1$, ($16'$) allows $r^2(t, \hat{z}) = 0$, i.e. describing infinitely many spherical membranes (centered at heights $z_n = 2n\omega$, and extending over $z$-intervals of length $L_t < 2\omega$) that at $t^2 = 1$ merge\footnote{near the points of merging one has $(sl(z))^2 = (cl(\varepsilon))^2 = (1-\varepsilon^2 + \frac{\varepsilon^4}{2}+ \ldots)^2 \approx 1-2\varepsilon^2$, i.e. $r^2(t, \varepsilon) = (t^2 - 1)+2\varepsilon^2$, giving the singular shape $r(\varepsilon)= \sqrt{2}|\varepsilon|$ at $t = \pm 1$}  into one infinitely extended axially symmetric surface, topologically for all $t^2 >1$ a (growing for $t>1$, shrinking for $t < -1$) cylinder, smooth, and for $t^2 \rightarrow + \infty$
the contour becoming more and more flat $(r \rightarrow |t| \sqrt{1+\frac{sl^2(z)}{t^2}} \sim |t| + \frac{sl^2(z)}{2|t|})$; the points of minimal distance to the $z$-axis moving light-like, while all other points moving (outward for $t>0$, inward for $t<0$) with smaller velocity.\\
Finally, ($14''$) gives
\[
u(x^{\mu}) = -t^2+x^2+y^2+H^2(z) \tag{$15''$}
\]
\[
r^2(t,z) = t^2 - H^2(z), \tag{$16''$}
\]
\begin{align*}
(\frac{1}{2}\partial_{\mu}u)^2  & = t^2 - r^2 - H^2H'^2  \tag{$17''$} \\ 
    & = t^2-r^2-H^2-H^6 \\
    &  \cong -H^6 \leqslant 0.
\end{align*}
Taking $H (=slh)$ to be infinite at even multiples of its half-period $\tilde{\omega}$, this solution, in contrast to the string case, where one got one {\it single} growing closed curve, describes infinitely many spherical membranes that for $t>0$ grow (for $t<0$ shrink) but each being confined to `its' (finite) maximal $z$-range, $z_n$ between $2n\tilde{\omega}$ and $2(n+1)\tilde{\omega}$; their shape becomes flat(ter) as $t^2\rightarrow \infty$ (the radially outmost points, at heights $\hat{z}_n = (2n+1)\tilde{\omega}$, move light-like, all others slower).\\
Finally, I would like to discuss in detail a phenomena that is best illustrated with the minimal surface $\mathfrak{M}_2$ given by (s.a)
\begin{equation}\label{eq18} 
t^2 = x^2 + \sinh^2 z \;(=(t_{\pm}(x,z)^2),
\end{equation}
describing a (for $t>0$ steadily growing) convex smooth closed (time-dependent) curve $C_t$; note that $t_+(x,z) \geqslant 0$, defined on all of $\mathbb{R}^2$, is the time at which the curve passes the point $(x,z) \in \mathbb{R}^2$. The time-evolution of $C_t$ seems to contradict the (not particularly well known, but still) fact (independently found in \cite{PRS83}\cite{JH95}; see also \cite{KT82}\cite{EH09}, and for the possible
effects of singularities in the context of cosmic strings e.g.\cite{VSD}) 
that closed curves have to face a singularity in finite time, resp. that topologically cylindrical (time-like) minimal surfaces in $\mathbb{R}^{1,2}$ do not exist \cite{NT12} (some results on non-cylindrical ones have been obtained in \cite{P19}). 
\\How can (\ref{eq18}) circumvent the above singularity (-formation) theorem?\\
Clearly it must have to do with the fact that the two $z=0$ points on $C_t$ move with velocity 1, i.e. $\mathfrak{M}_2$ (otherwise being time-like) containing two light-like lines. But what exactly can be learnt from the above example? (as we will see, the answer is astonishingly rich).\\
String theorists commonly assume the existence of coordinates on the world-sheet in which the metric is conformally flat, and the motion is usually described by 
\begin{equation}\label{eq19} 
x^{\mu} = 
\begin{pmatrix}
t \\ \vec{x}(t,\varphi)
\end{pmatrix}
\end{equation}
where 
\begin{equation}\label{eq20} 
\dot{\vec{x}}\vec{x}' = 0, \qquad \dot{\vec{x}}^2 + \vec{x}'^2 = 1. 
\end{equation}
One way to arrive at (\ref{eq20}) is to notice that if choosing an orthogonal parametrization (i.e. $\dot{\vec{x}}\vec{x}'$ = 0, the string moving for each $\varphi$ orthogonal to itself) the equations of motion \\
\begin{equation}\label{eq21} 
\frac{1}{\sqrt{G}}\partial_{\alpha}(\sqrt{G}G^{\alpha\beta}\partial_{\beta}x^{\mu}) = 0
\end{equation}
for $\mu = 0$ imply
\begin{equation}\label{eq22} 
\frac{\partial}{\partial t}\left( \rho := \sqrt{\frac{\vec{x}'^2}{1-\dot{\vec{x}}^2}} \right)  = 0
\end{equation}
and, given (\ref{eq22}), the rest of (\ref{eq21}) becomes
\begin{equation}\label{eq23} 
\ddot{\vec{x}} = \frac{1}{\rho}\big( \frac{1}{\rho} \vec{x}' \big)'
\end{equation}
(where $'$ denotes differentiation with respect to $\varphi$, and, as usual, $\cdot$ differentiation with respect to time). Folklore says that starting with the Schild action \cite{S77} (see also \cite{H96}) allows to incorporate light-like parts of $\mathfrak{M}_2$, and that $\rho$ can be put $=1$; indeed $\underset{\sim}{\vec{x}}(t,\tilde{\varphi}) = \vec{x}(t,\varphi)$ with $\frac{d\tilde{\varphi}}{d \varphi} = \rho(\varphi)$ is in principle a correct argument for putting $\rho(\varphi)$ to a constant, but (not always spelled out) if one wants $\tilde{\varphi}$ to have the same range than $\varphi$ (e.g. $[0,2\pi]$) one can {\it not} put $\rho = 1$, but only equal to 
\begin{equation}\label{eq24} 
\rho_0 := \int \rho(\varphi)d(\varphi), 
\end{equation}
which however (tacitly assuming that $\rho_0$ is finite) can in principle, in an additional step, be absorbed in a re-scaling of time, hence finally obtaining 
\begin{equation}\label{eq25} 
\square \vec{x} := (\partial_t^2 - \partial_{\varphi}^2)\vec{x} = 0;
\end{equation}  
which is not only consistent with (\ref{eq20}) ($\vec{x} \in \mathbb{R}^{D-1}$) but in co-dimension 1 (for strings: $D=3$; for M(em)branes: $D=M+2$, see e.g. \cite{JH95} for analogous reasonings) {\it implied} by (\ref{eq20}), as long as $\dot{\vec{x}}$ and $\vec{x}'$ are linearly independent.\\
In any case, one then can take
\begin{equation}\label{eq26} 
\begin{split}
\dot{\vec{x}} & = -\sin (f-g)
\begin{pmatrix}
\cos(f+g) \\ \sin(f+g)
\end{pmatrix}\\
\vec{x}' & = \cos(f-g)
\begin{pmatrix}
-\sin(f+g) \\ \cos(f+g)
\end{pmatrix}
\end{split}
\end{equation}  
where $f = f(\varphi+t =: \varphi_+)$ and $g = g(\varphi_- := \varphi -t)$; written in this form, the curvature of $C_t$ follows to be \cite{JH95}
\begin{equation}\label{eq27} 
\kappa = \frac{f' + g'}{\cos (f-g)}.
\end{equation} 
A crucial part of (\ref{eq27}) implying singularity formation in finite time is to use that the curve is closed.\\
To see what goes wrong when taking the above mentioned example, one can try to parametrize (\ref{eq18}) according to (\ref{eq20}). As difficult as such `parametrization-problems' generally are, one {\it can} actually succeed by first noticing {\it another} (conformally flat) parametrization, namely
\begin{equation}\label{eq28} 
\underset{\sim}{x}^{\mu}(u,v) = 
\begin{pmatrix}
\sinh u  \cosh v \\ \sinh u \sinh v \\ u
\end{pmatrix}
\end{equation} 
giving $\tilde{G}_{\alpha\beta} = \sinh^2 u 
\big( \begin{smallmatrix}
1 & 0 \\ 0 & -1
\end{smallmatrix} \big)
$, and the components of (\ref{eq28}) obviously being annihilated by $\partial^2_u - \partial^2_v$. To find $\varphi(u,v)$ one can use 
\begin{equation*}
\begin{split}
t(u,v) & = \sinh u \cosh v\\
t_u & = \cosh u \cosh v \\
t_v & = \sinh u \sinh v,
\end{split}
\end{equation*} 
and the general law of how the metric tensor transforms under reparametrizations,
\begin{equation}\label{eq29} 
\begin{split}
\tilde{G}_{\alpha\beta} & = (J^T)_{\alpha} \;^{\alpha'}G_{\alpha'\beta'} J^{\beta'}\,_{\beta}\\
(J) & = \big( \frac{\partial t \varphi}{\partial u v} \big) = \binom{t_u t_v}{\varphi_u \varphi_v}\\
(G) & =  
\begin{pmatrix}
1 - \dot{\vec{x}}^2 & 0 \\ 0 & -\vec{x}'^2
\end{pmatrix} \sim
\begin{pmatrix}
1 & 0 \\ 0 & -1
\end{pmatrix},
\end{split}
\end{equation} 
which together implies the PDE
\begin{equation}\label{eq30} 
\frac{\varphi_u \varphi_v}{\varphi_u^2 + \varphi_v^2} = \frac{t_u t_v}{t^2_u + t^2_v}
\end{equation}
for $\varphi$.
\begin{equation}\label{eq31} 
\varphi(u,v) = \cosh u \sinh v
\end{equation}
not only solves (\ref{eq30}), but (using elementary addition theorems for the hyperbolic functions) allows to explicitly invert,
\begin{equation}\label{eq32} 
\begin{split}
u+v & = \text{arcsinh} (\varphi + t)\\
v-u & = \text{arcsinh} (\varphi- t).
\end{split}
\end{equation}
It then follows that
\begin{equation}\label{eq33} 
\vec{x}(t,\varphi) = \frac{1}{2}\binom{\sqrt{\varphi_+^2 +1}-\sqrt{\varphi_-^2 +1}}{\text{arcsinh} \varphi_+ - \text{arcsinh} \varphi_-},
\end{equation}\\[0.15cm]
\begin{equation}\label{eq34} 
\begin{split}
2\dot{\vec{x}} & = \frac{1}{\sqrt{\varphi^2_+ +1}}\binom{\varphi_+}{1} + \frac{1}{\sqrt{\varphi^2_- +1}}\binom{\varphi_-}{1}
 =: \vec{v}_+ + \vec{v}_-\\[0.15cm]
 2\vec{x}' & = \vec{v}_+ - \vec{v}_-,
\end{split}
\end{equation}
trivially allowing to verify that the $so$ parametrized solutions do satisfy (\ref{eq20}) (and (\ref{eq18})!)
As nice as this is, one should point out the following three (related) problems:\\
Firstly, (\ref{eq33}) describes only {\it half} of $\mathfrak{M}_2$ (stemming from the first and third component in (\ref{eq28}) necessarily having the same sign, while (\ref{eq18}) allows both positive and negative $z$ for each given $t$).\\
Secondly, $\varphi \in \mathbb{R}$ (cp.(\ref{eq31})), i.e. {\it not} confined to a compact interval (like $[0,2\pi]$).\\
Thirdly, if one was to calculate the energy\footnote{I thank G. Huisken for very stimulating discussions related to this point.} of the solution, one would find that it diverges. As in a Hamitonian 2+1 formulation (analogously for M(em)branes \cite{H93}) $\rho^2 := \frac{\vec{x}'^2}{1-\dot{\vec{x}}^2}$ is nothing but the square of the energy density, the above means that even if one resolved the first two problems, the third one would remain, i.e. 
even if one found a nice {\it orthogonal} parametrization of the complete string with $\tilde{\varphi} \in [0,2\pi]$, $\rho_0 = \int \tilde{\rho}(\tilde{\varphi})\thinspace d\tilde{\varphi}$ would {\it not} be finite, hence a priori {\it not} allowing to go from (\ref{eq23}) to (\ref{eq25}). Physically, the non-finiteness of the energy is of course clear from the beginning, as any finite energy closed string could not grow forever. Mathematically, the two $z = 0$ points on $C_t$ (corresponding to $\varphi \to \pm \infty$) are `difficult to pass'.\\
One can pinpoint how (\ref{eq18}) evades the singularity formation theorem by explicitely calculating $f$ and $g$ appearing in (\ref{eq27}); one finds
\begin{equation}\label{eq35} 
\begin{split}
f = \frac{1}{2} \arctan(\varphi_+ )+ \frac{\pi}{4}\\
g = \frac{1}{2} \arctan(\varphi_-) - \frac{\pi}{4}
\end{split}
\end{equation}
which indeed has the property that $\cos(f-g)$ vanishes only for $\varphi_+ = \varphi_-$, i.e. $t=0$. Having obtained the solution in this form, it is easy to see that replacing $\arctan$ by any strictly monotonic\footnote{one may of course also relax this condition (while loosing some geometric properties).} function $h$ going to $\pm \frac{\pi}{2}$ at infinity will provide closed string solutions that (in one t-direction) are growing (forever), and contain 2 points $(z = 0)$ moving with unit velocity; obtaining (apart from containing 2 light-like lines) time-like minimal hypersurfaces $\mathfrak{M}_2 \subset \mathbb{R}^{1,2}$ that are reflection symmetric in all 3 directions and having only one single singularity (the origin, where the loop degenerates to a point). Just as with (\ref{eq35}),
\begin{equation}\label{eq37} 
\begin{split}
f(\varphi_+) = \frac{1}{2} h(\varphi_+) + \frac{\pi}{4}\\
g(\varphi_-) = \frac{1}{2} h(\varphi_-) - \frac{\pi}{4}
\end{split}
\end{equation}
strictly speaking parametrizes only half of $\mathfrak{M}_2$, but due to
\begin{equation}
\begin{split}
\frac{\partial z(t,z)}{\partial x}\bigg|_{x = x(t,\varphi)} &= \frac{\cos(h(\varphi_+))-\cos(h(\varphi_-))}{\sin(h(\varphi_+))-\sin(h(\varphi_-))}\\[0.15cm]
 &= -\tan\left(\frac{h(\varphi_+)-h(\varphi_-)}{2}\right) ,
\end{split}
\end{equation}
which follows from
\begin{align*}
\begin{pmatrix}
1 & 0\\
\dot{\varphi} & \varphi' 
\end{pmatrix} = \bigg( \frac{\partial t\varphi}{\partial tx} \bigg) =\bigg( \frac{\partial tx}{\partial t\varphi}\bigg)^{-1} = \begin{pmatrix}
1 & 0\\
\dot{x} & x' 
\end{pmatrix}^{-1} = \begin{pmatrix}
1 & 0\\
\frac{-\dot{x}}{x'} & \frac{1}{x'} 
\end{pmatrix} ,
\end{align*}
and $z(t,x) = y(t,\varphi(t,x))$ implying $z_x = y'\varphi_x = y'/x'$, all the solutions defined by (36)
have the property that $z_x|_{x = x(t,\varphi)}$ is a monotonic function of $\varphi \in (-\infty, +\infty)$ that diverges at both ends, hence (due to the $z$ going to $-z$ symmetry of the original equations) allowing to define (the time-evolution of) smooth (convex) closed curves $C_t^{(h)}$ for any (above specified) choice of $h$; hence showing that the world of infinite-energy string-solutions is not really `smaller' than that of conventional string solutions. Thus substantially enlarging `String Theory'.\\
While it is good to have the above examples where everything can be spelled out in elementary, explicit, terms the same reasoning could equally well be applied to the hyperbolic membrane solution (\ref{eq15}),
\begin{equation}\label{eq38} 
\begin{split}
t^2 = x^2 + y^2 + H^2(z)\\
H'^2 = 1 + H^4
\end{split}
\end{equation}
which (at least `one half', s.a.) could (first) be parametrized by
\begin{equation}\label{eq39} 
\tilde{x}^\mu(u,v,w) = \begin{pmatrix}
H(u)\cosh(v)\\
H(u)\sinh(v)\cos(w)\\
H(u)\sinh(v)\sin(w)\\
u 
\end{pmatrix}
\end{equation}
and then as
\begin{equation}\label{eq40} 
x^\mu(t,\varphi,\psi) =\begin{pmatrix}
t\\
r(t,\varphi)\cos(\psi)\\
r(t,\varphi)\sin(\psi)\\
z(t,\varphi)
\end{pmatrix}
\end{equation}
\begin{equation}\label{eq41}
\begin{split}
\begin{pmatrix}
H'^2-1=H^4 & 0 & 0\\
0 & -H^2 & 0\\
0 & 0 & -H^2\sinh^2(v)
\end{pmatrix} = \tilde{G}_{\alpha,\beta}\\[0.15cm]
= J^T \begin{pmatrix}
1-\dot{r}^2-\dot{z}^2 & 0 & 0\\
0 & -r'^2-z'^2 & 0\\
0 & 0 & -r^2
\end{pmatrix}J\\[0.15cm]
J = \left(\frac{\partial t\varphi\psi}{\partial uvw}\right)
\end{split}
\end{equation}
with $\psi(u,v,w)=w$ and
\begin{equation}\label{eq42} 
t(u,v,w)=H(u)\cosh(v),\thickspace t_u=H'\cosh(v),\thickspace t_v=H\sinh(v)
\end{equation}
giving again a single PDE for $\varphi = \varphi(u,v)$, this time reading
\begin{equation}\label{eq43} 
\frac{\varphi_u\varphi_v}{H^2\varphi_u^2+\varphi_v^2H^4}=\frac{HH'\sinh(v)\cosh(v)}{H^6\sinh^2(v)+H^2H'^2\cosh^2(v)}.
\end{equation}
Although the rhs is more complicated than in the string case one can (as presumably $\it always$ for axially symmetric M-brane solutions for which t is a simple product) still find $\varphi$ more or less explicitly:
Making the Ansatz $\varphi(u,v)=h(H(u,v))\sinh(v)$ one obtains $h'(1+H^4)=H^3h$, hence 
\begin{equation}\label{eq44}
\varphi(u,v) = (1+H^4(u))^\frac{1}{4}\sinh(v)  ,
\end{equation}
together with (\ref{eq42}) implying e.g. (where $H(u)\neq 0$, resp. $v\neq 0$)
\begin{equation}\label{eq45}
\frac{\varphi^4}{1+H^4}=\left(\frac{t^2}{H^2}-1\right)^2,\thickspace \frac{-t^4}{(\cosh(v))^4}+\frac{\varphi^4}{(\sinh(v))^4}=1
\end{equation}
as well as
\begin{equation}\label{eq46}
\rho^2:=\frac{r^2\vec{x}'^2}{1-\dot{\vec{x}}^2}=r^2\left(\frac{-G_{11}}{G_{00}}\right)=r^2\left(\frac{H^2t_u^2+H^4t_v^2}{H^2\varphi_u^2+H^4\varphi_v^2}\right)=\varphi^2.
\end{equation}
To double check, one may express $\dot{u}=\dot{z}$, $u'=z'$,$\dot{v}$ and $v'$ as functions of $u$ and $v$ by differentiating (\ref{eq42}) and (\ref{eq44}) with respect to $t$ and $\varphi$, and then, with $\dot{r}=\dot{u}H'\sinh+H(\cosh)\dot{v}$ and $r'=u'H'\sinh+\cosh Hv'$, verify that the equations of motion for axially symmetric membranes (cp.\cite{JH19},\cite{EHHS15}
and references therein)
\begin{equation}\label{eq47}
\dot{r}r' + \dot{z}z' = 0, \thickspace \dot{r}^2+\dot{z}^2+r^2\frac{(r'^2+z'^2)}{\rho^2}=1
\end{equation}
are indeed satisfied.

\end{document}